\documentclass{webofc}
\usepackage[varg]{txfonts}   
\usepackage{listings}
\usepackage{hyperref}
\begin{document}
\title{OGLE Cepheids and RR Lyrae Stars in the Milky Way}

\author{\firstname{Andrzej} \lastname{Udalski}\inst{1}\fnsep\thanks{\href{mailto:udalski@astrouw.edu.pl}{\tt udalski@astrouw.edu.pl}}
on behalf of the OGLE survey team
}

\institute{Warsaw University Observatory, Al.~Ujazdowskie 4, 00-478
Warszawa, Poland}

\abstract{
We present new large samples of Galactic Cepheids and RR Lyrae stars
from the OGLE Galaxy Variability Survey.
}
\maketitle
\vspace*{-10pt}
\section{Introduction}\label{sec:intro}

The Optical Gravitational Lensing Experiment (OGLE) is one of the
largest sky variability surveys. In its fourth phase it regularly
monitors over a billion stars. The main OGLE observing targets are the
densest stellar regions of the sky: the Galactic center and Magellanic
Clouds \cite{uda2015}.

One of the most important outcome of the project is the {\it OGLE
Collection of Variable Stars} (OCVS). It currently consists of almost a
million well characterized periodic variable stars of many different
types \cite{sos2016b}. The OCVS contains a full variety of pulsating
stars from the main astrophysical ``laboratories'' like the Magellanic
System and the Galactic bulge. For example, the current sample of RR
Lyrae stars counts over 38000 objects in the Galactic bulge  and over
45000 stars in the Magellanic System (Magellanic Clouds and Magellanic
Bridge) \cite{sos2014,sos2016a}. The number of Cepheids in the
Magellanic System reaches 10\ 000 pulsators \cite{sos2015}. The OGLE
samples are very pure and complete. They were used for many interesting
projects related, among others, to investigations of the structure of
the Galactic center \cite{pie2015} and Magellanic System
\cite{jac2016,jac2017}.

In 2013 a new OGLE long-term survey -- OGLE Galaxy Variability Survey
(OGLE GVS) -- was initiated in the frame of the OGLE-IV phase. An area
of about 2000 square degrees in the sky is regularly monitored. The
survey covers a strip of $\pm 3$ degree width along the Galactic plane
for all galactic longitudes accessible from the OGLE observing site --
Las Campanas Observatory, Chile. Additionally, a large area of the
external Galactic bulge is also covered.

After collecting large sample of photometric data (over 100 epochs),
spanning three observing seasons, the region of the Galactic disk with
$190<l<345$~degrees is ready for exploration. Here we present the
preliminary report on the first quick search for Cepheids and RR Lyrae
stars in these fields of the OGLE GVS.  

\vspace*{-6pt}
\section{Observations}\label{sec:observations}

The OGLE-IV survey has been conducted using the 1.3-m Warsaw telescope
with a 32 CCD detector wide field mosaic camera located at the Las
Campanas Observatory, Chile. The field of view of the OGLE-IV camera is
1.4 square degrees with the pixel scale of 0.26 arsec/pixel. The OGLE
GVS has been started in February 2013 and it is still conducted albeit
with different cadence and priority of fields compared to previous
seasons. The main variability survey is a shallow one with the exposure
time of 25 seconds. Most observations are collected in the {\it I}-band
filter for variability search. 120--150 epochs in each of the fields
presented here have been collected so far. Also, about ten or more
observations have been secured in the {\it V}-filter for color
information.

The collected images have been reduced with the OGLE standard
photometric pipeline which is based on the image subtraction technique
-- DIA \cite{uda2015}. The faint limit for variability search in the
OGLE GVS is about $I\approx 18.5$~mag.

\vspace*{-6pt}
\section{Cepheids and RR Lyrae Stars from the OGLE GVS}\label{sec:ceps}

We carried out the pilot search for pulsating variables of
Cepheid and RR Lyrae types among about 560 million stellar sources
detected on the OGLE GVS images of the Galactic plane. This ``fast
search'' is different than the typical global OGLE variable stars
searches. It has been conducted on pre-selected subsample of stars --
only those which displayed clear variability signatures on the
difference images. Minimum 15 detections of variable flux in the
difference images were required to include a star to this sample. 

The pre-selected variable star candidates were then subject to period
search. Two techniques were used: analysis of variance, {\sc AoV}
\cite{asc1989}, and Fourier analysis implemented as {\sc fnpeaks}
program \cite{kol2016}. Light curves of objects with significant
periodic signals were further investigated. First, their shape was
fitted with Fourier series and the Fourier parameters $R_{21}$ and
$\phi_{21}$ were calculated. In the next step of preselection, only
candidates with Fourier parameters in a wide range covering typical
regions occupied by Cepheids and RR Lyrae stars on the $R_{21}$ vs. $log
P$ and $\phi_{21}$ vs. $log P$ diagrams were left. Finally, the
candidates were visually inspected and objects with light curves of
non-pulsating origin (e.g., eclipsing binaries) were removed from the
ultimate list of Cepheid/RR Lyrae candidates.

\begin{figure*}
\centering
\includegraphics[width=12cm,clip]{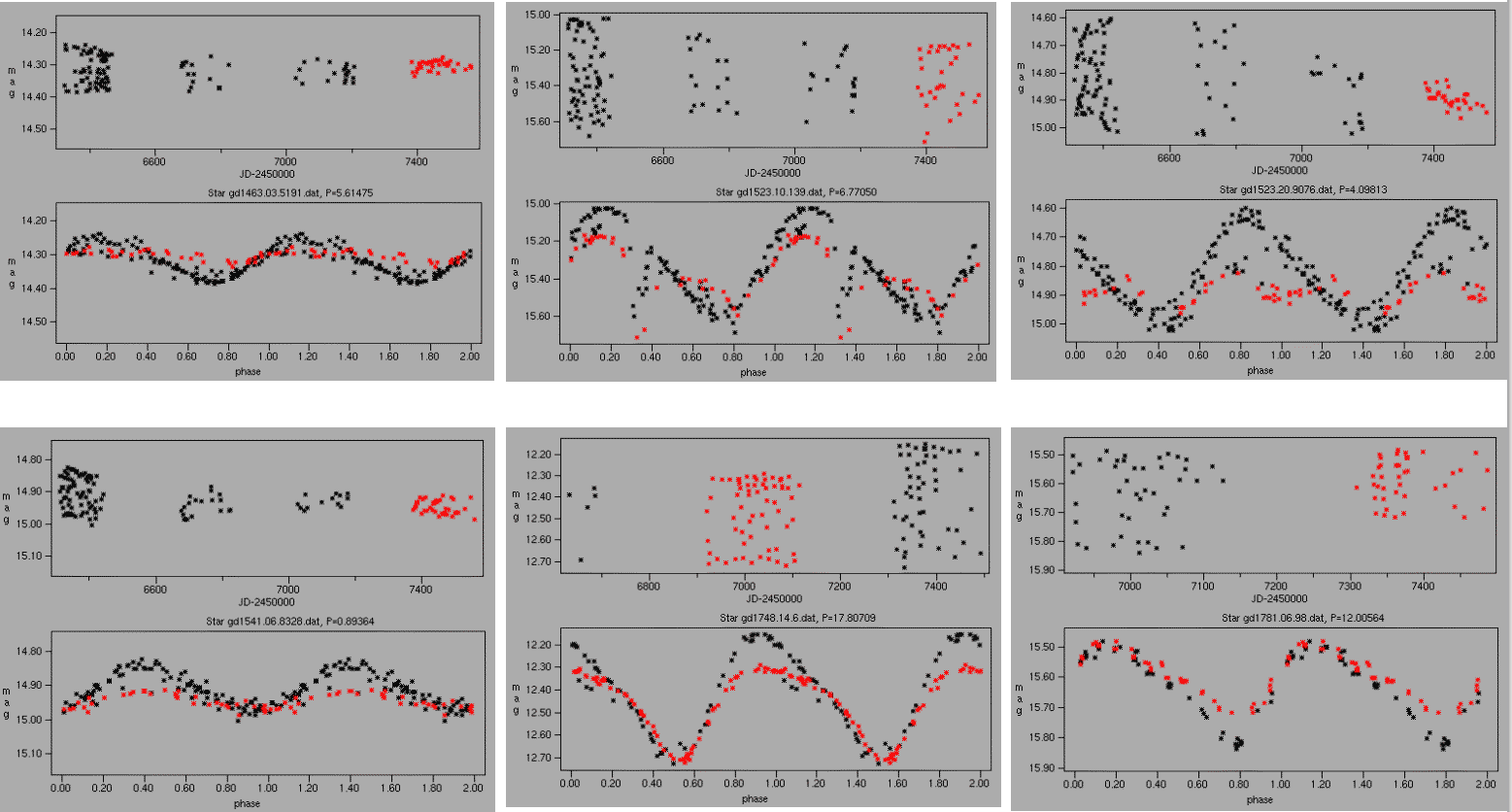}
\caption{Example light curves of variable stars mimicking pulsating-like
variability behavior. Red points mark observing season with considerably
different shape of the light curve.}
\label{fig:ogle-fig-1}       
\end{figure*}

Surprisingly, the list of objects at this stage was still severely
contaminated. The main source of fake identifications were spotted stars
which can mimic very well the light curve of pulsating star. On the
other hand the shape of the variability in such stars evolves with time
and in the case of long-term monitoring, as during OGLE GVS, spotted
stars can be extracted and separated from genuine pulsators.
Figure~\ref{fig:ogle-fig-1} presents light curves of a few examples of
spotted stars mimicking in some observing seasons pulsating stars.

\begin{figure*}
\centering
\includegraphics[width=12cm,clip]{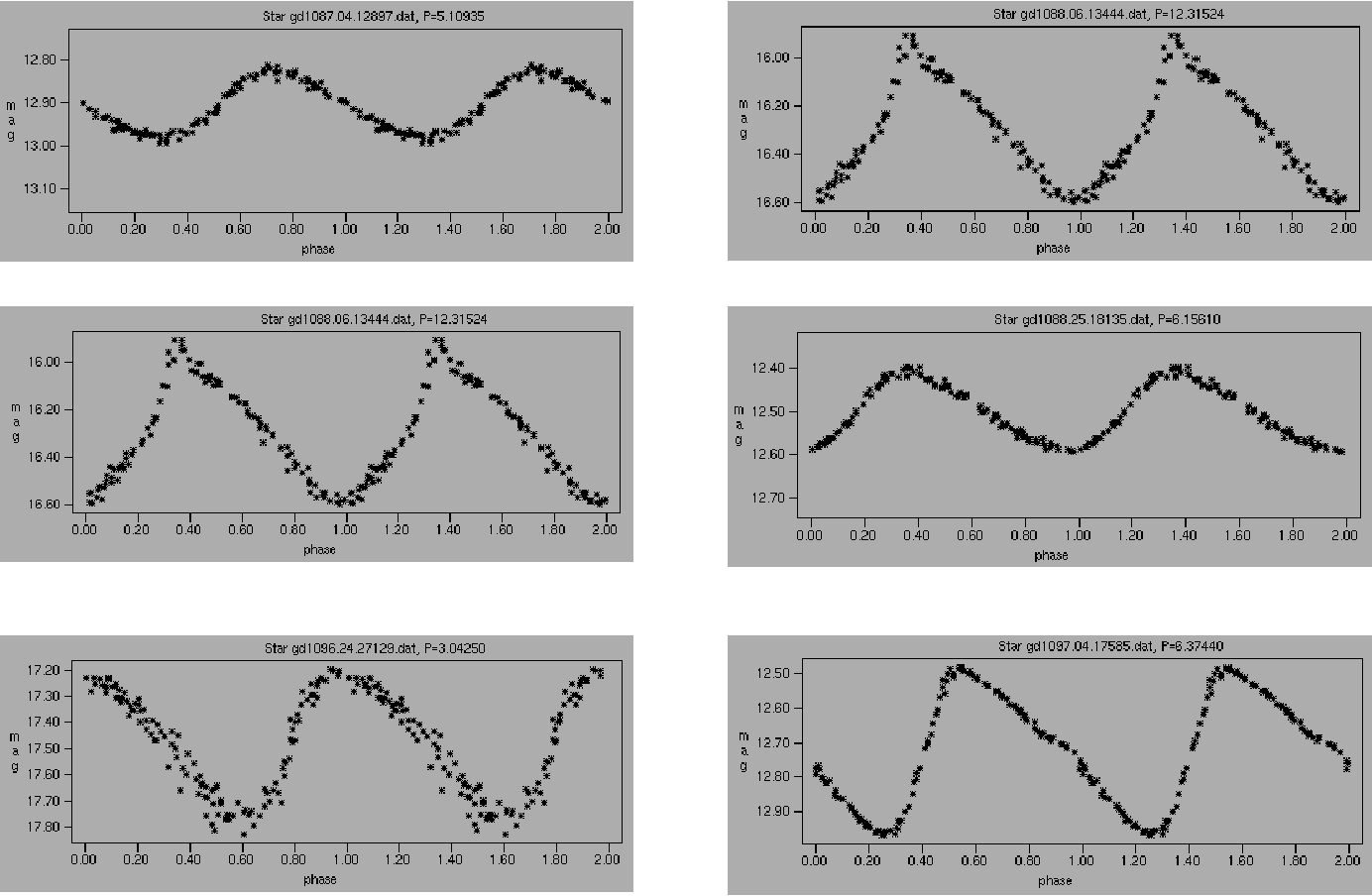}
\caption{Example light curves of newly discovered Galactic disk
Cepheids.}
\label{fig:ogle-fig-2}       
\end{figure*}

After visual inspection of the OGLE Cepheid/RR Lyrae candidates 1930 and
2170 objects were left as very likely Galactic Cepheids and RR Lyrae
type stars, respectively. At this stage of analysis the boundary between
these two groups was set artificially at $P=0.7$~day.
Figure~\ref{fig:ogle-fig-2} presents a small gallery of newly discovered
Galactic Cepheids. The list of Cepheids contains classical Cepheids as
well as type II objects.

Preliminary lists of the OGLE Cepheid and RR Lyrae candidates will
additionally be refined when the color data are available. Nevertheless,
the purity of the sample is already high.  The completeness of these
samples is different. While the Galactic Cepheids are bright enough to
be detected on the OGLE subtracted images, fainter RR Lyrae stars
located farther in the halo are certainly missed in our preselected
sample of analyzed stars. We expect that many additional fainter RR
Lyrae stars will be detected during the final exensive search for
variables in the OGLE GVS fields.

\begin{figure*}
\centering
\includegraphics[width=11cm,clip]{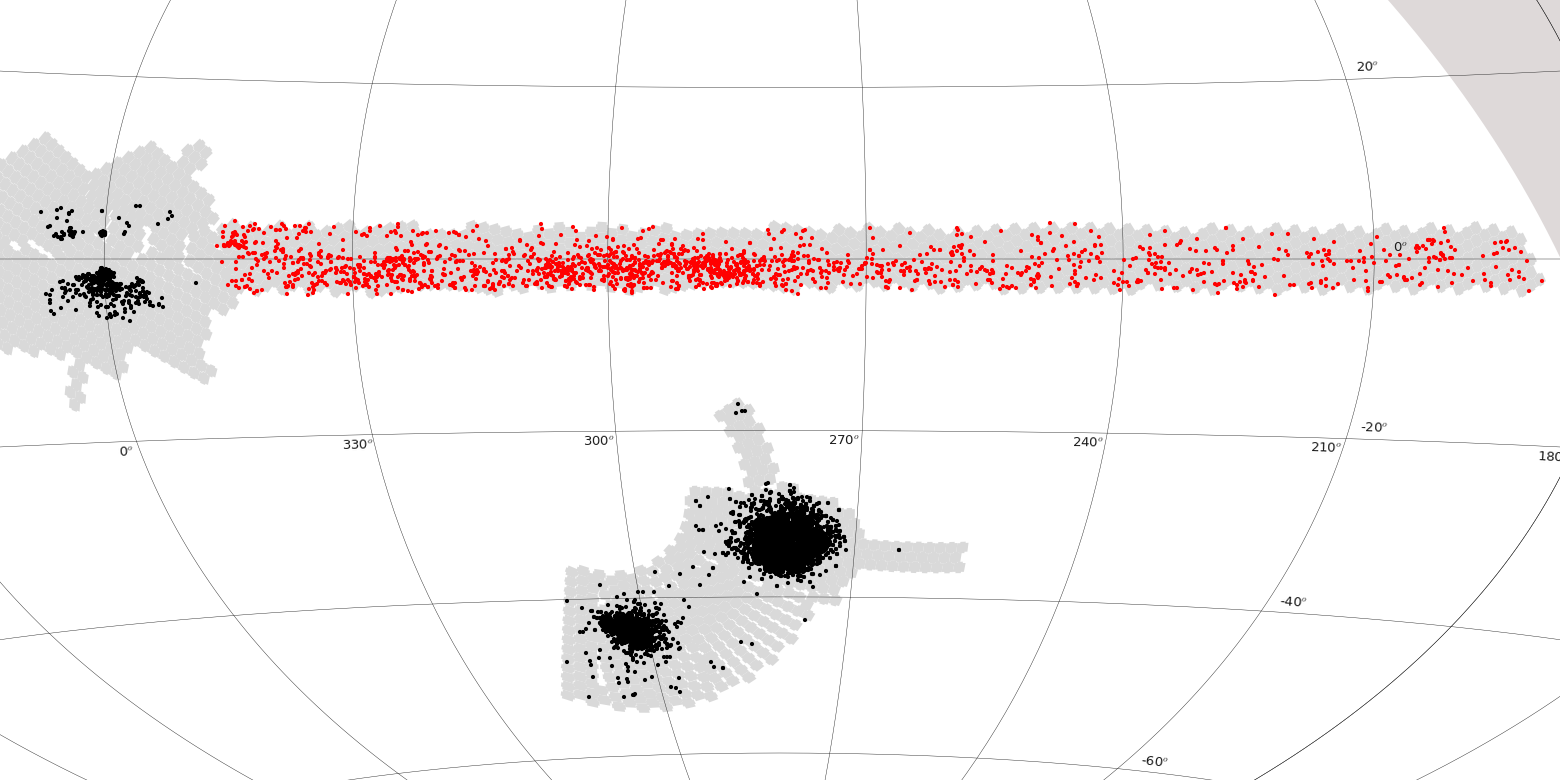}
\caption{OGLE Cepheids in the sky. The sky map is in the Galactic
coordinates ($l,b$). Light gray shaded areas show the regions regularly
monitored by the OGLE-IV survey. Dark gray shaded region in the upper
right corner indicates part of the sky not accessible from Las Campanas
Observatory. Red dots present Cepheid star candidates found in the OGLE
GVS fields. Black dots mark positions of the OGLE Cepheids in other
OGLE-IV targets.}
\label{fig:ogle-fig-3}       
\end{figure*}

\begin{figure*}
\centering
\includegraphics[width=11cm,clip]{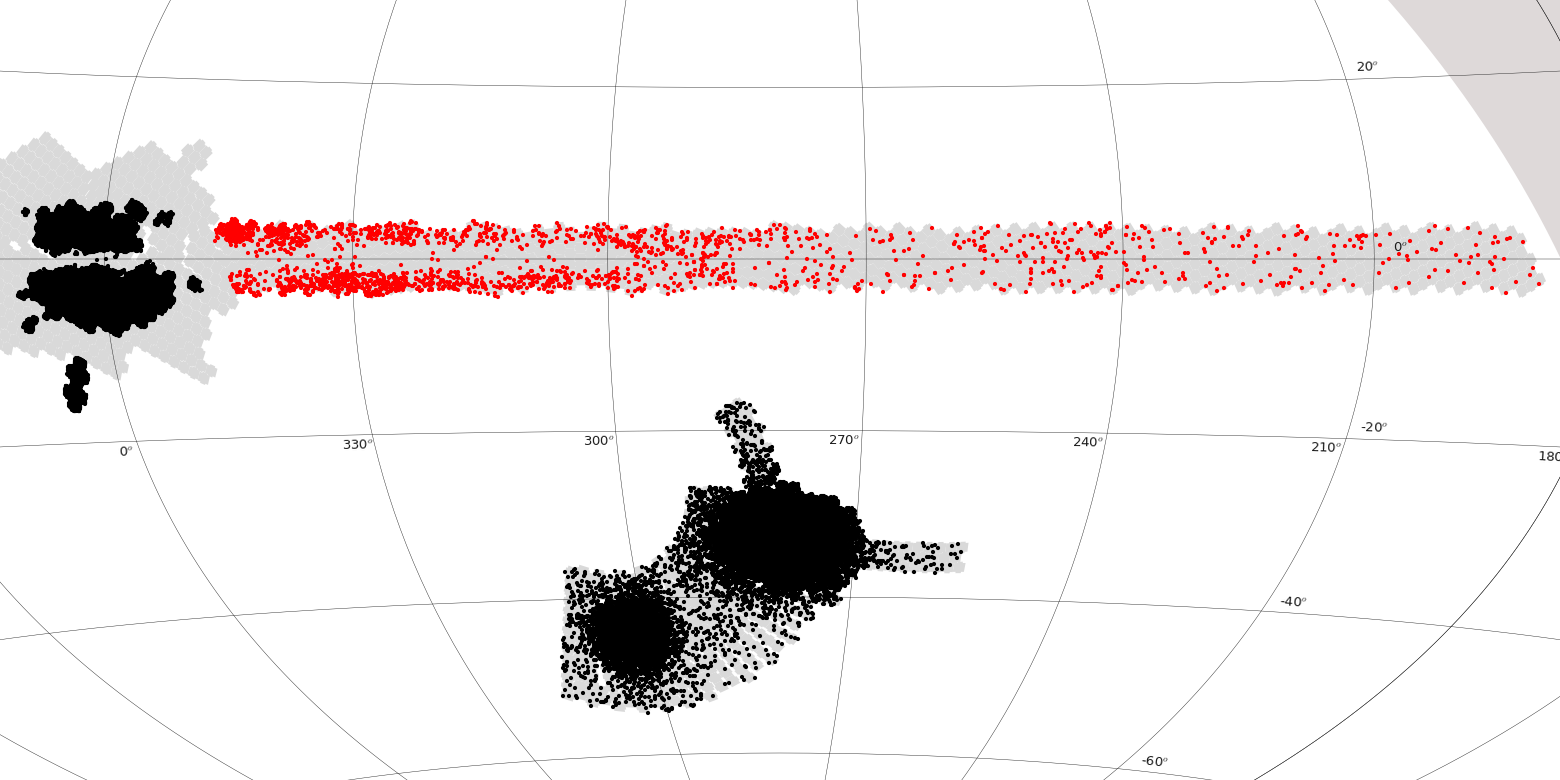}
\caption{OGLE RR Lyrae stars in the sky. The sky map is in the          
Galactic coordinates ($l,b$). Light gray shaded areas show the regions 
regularly monitored by the OGLE-IV survey. Dark gray shaded region in
the upper right corner indicates part of the sky not accessible from Las
Campanas Observatory.  Red dots present RR Lyrae star candidates found
in the OGLE GVS fields. Black dots mark positions of the OGLE RR Lyrae
stars in other OGLE targets.}
\label{fig:ogle-fig-4}       
\end{figure*}

\begin{figure*}
\centering
\includegraphics[width=14.0cm,clip]{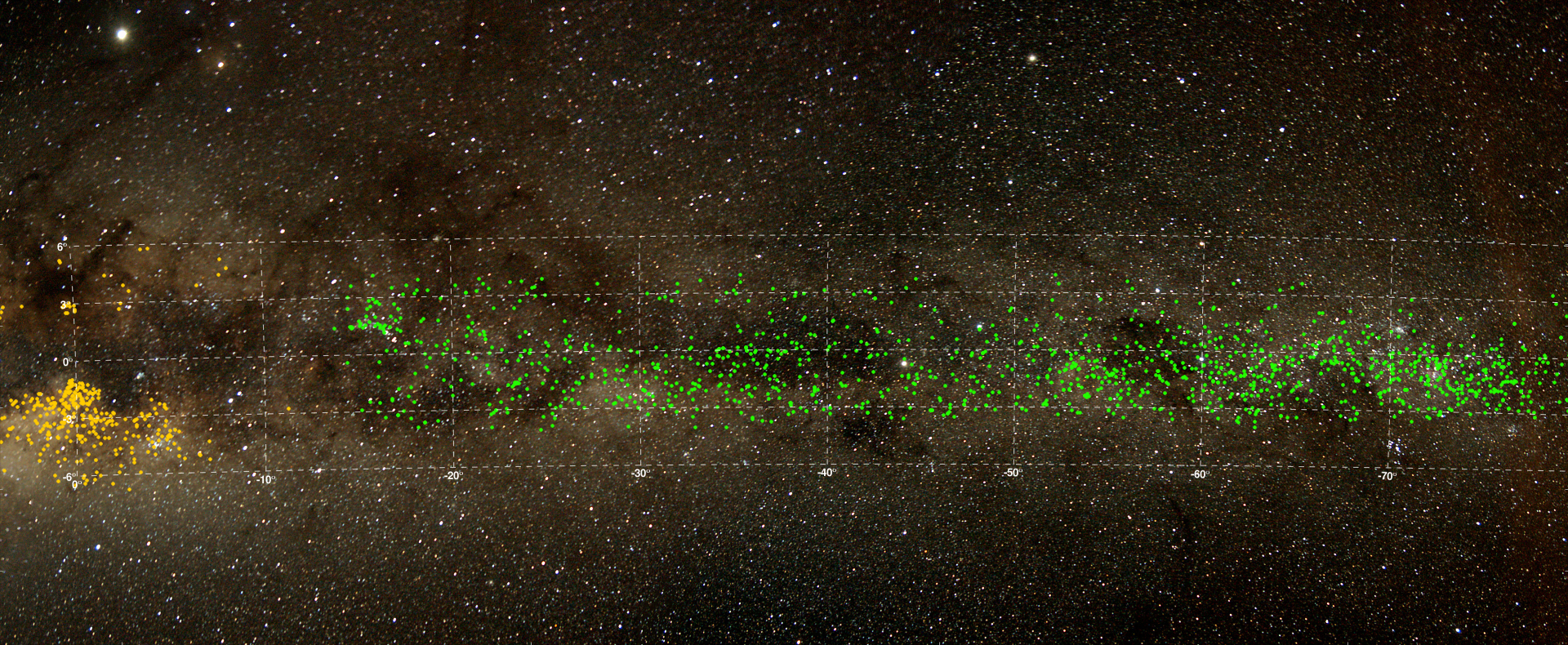}

\caption{Details of the distribution of OGLE Cepheids in the Galactic
plane. Green dots indicate positions of the OGLE-IV Cepheid candidates
from the OGLE GVS. Yellow dots are OGLE Cepheids from the Galactic bulge
fields.}

\label{fig:ogle-fig-5}       
\end{figure*}

\begin{figure*}
\centering
\includegraphics[width=14.0cm,clip]{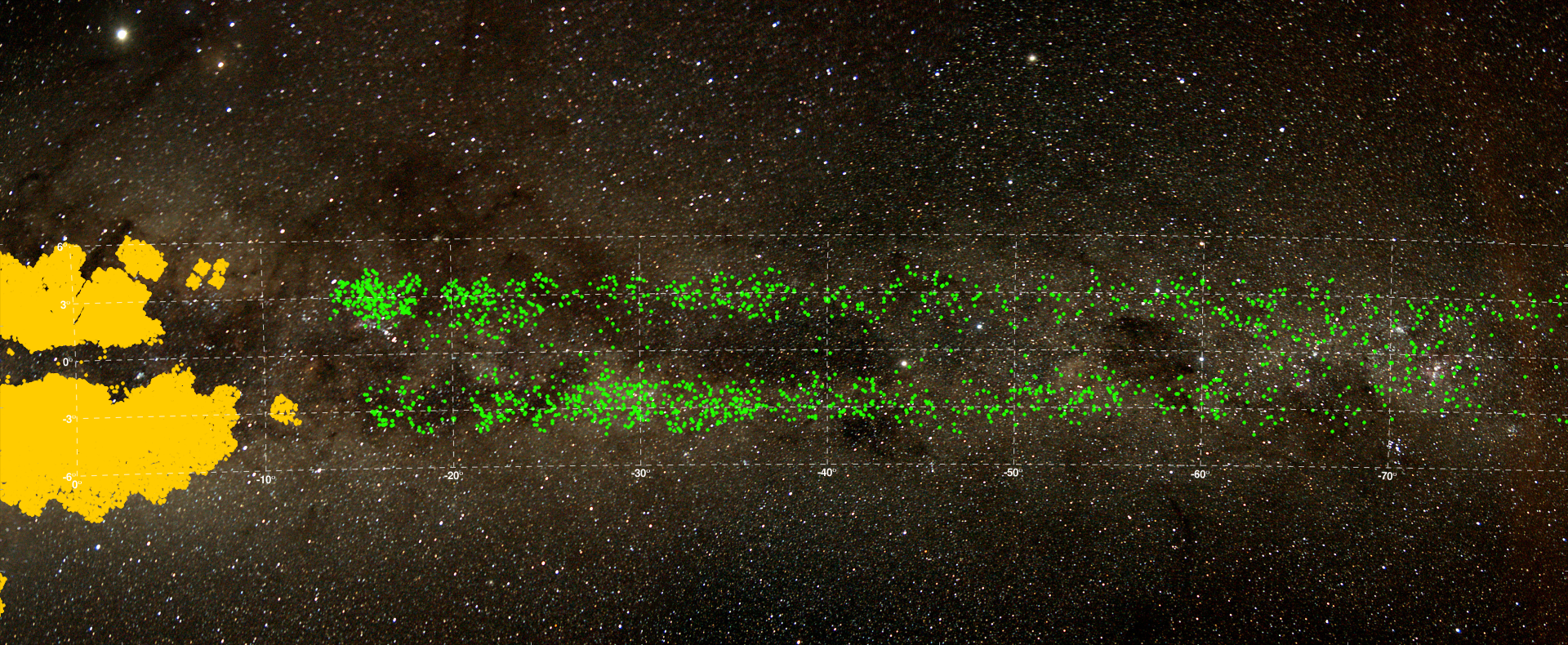}

\caption{Details of the distribution of OGLE RR Lyrae stars in the
Galactic plane. Green dots indicate positions of the OGLE-IV RR Lyrae
star candidates from the OGLE GVS. Yellow dots are OGLE RR Lyrae stars
from the Galactic bulge fields.}

\label{fig:ogle-fig-6}       
\end{figure*}

Figure~\ref{fig:ogle-fig-3} and \ref{fig:ogle-fig-4} show the
distribution of new OGLE Galactic Cepheids and RR Lyrae stars on the map
of the sky, respectively. Figure~\ref{fig:ogle-fig-5} and
\ref{fig:ogle-fig-6} present distribution of these groups of stars in
the real sky.    Even without deeper analysis one can easily notice
differences in distributions. For example, while RR Lyrae stars are
generally located symmetrically around the Galactic plane, Cepheids lie
mostly below the plane over a long range of observed galactic
longitudes.

The new OGLE samples of Galactic Cepheids and RR Lyrae stars will be
released to the community in the next months {\it via} the OCVS. They
will constitute the next huge OGLE datasets of these important pulsating
stars.  

\begin{acknowledgement} 
\noindent\vskip 0.2cm
\noindent {\em Acknowledgments}:
The OGLE project has received funding from the National Science Centre,
Poland, grant MAESTRO 2014/14/A/ST9/00121.
\end{acknowledgement}

%
%

\end{document}